\documentclass[10pt,letterpaper,compsoc,conference]{iiswc22}

\usepackage{cite}
\usepackage{amsmath,amssymb,amsfonts}
\usepackage{algorithmic}
\usepackage{graphicx}
\usepackage[dvipsnames]{xcolor}
\usepackage[final]{microtype}
\usepackage[italic]{mathastext}
\usepackage{libertine}
\usepackage[T1]{fontenc}
\usepackage{textcomp}
\usepackage[varqu,varl]{zi4}
\usepackage[all]{nowidow}
\usepackage[auth-lg,affil-it]{authblk}
\usepackage[keeplastbox]{flushend}

\usepackage[utf8]{inputenc}
\usepackage{caption}
\usepackage{subfig}
\usepackage{lineno,hyperref}
\usepackage{comment}
\usepackage{multirow}
\usepackage{makecell}
\usepackage{tabularx}

\begin{document}


\title{SpotLake: Diverse Spot Instance Dataset Archive Service}


\renewcommand\Authsep{\qquad}
\renewcommand\Authand{\qquad}
\renewcommand\Authands{\qquad}


\author{Sungjae Lee}
\author{Jaeil Hwang}
\author{Kyungyong Lee}
\affil{Department of Computer Science, Kookmin University, Seoul, South Korea\\
\{sungjae, jaeil, leeky\}@kookmin.ac.kr
}

\maketitle


\begin{abstract}
Public cloud service vendors provide a surplus of computing resources at a cheaper price as a spot instance. Despite the cheaper price, the spot instance can be forced to be shutdown at any moment whenever the surplus resources are in shortage. To enhance spot instance usage, vendors provide diverse spot instance datasets. Among them, the spot price information has been most widely used so far. However, the tendency toward barely changing spot price~\cite{spot-price-change-2017, spot-price-policy-change-2017-irwin} weakens the applicability of the spot price dataset. Besides the price dataset, the recently introduced spot instance availability and interruption ratio datasets can help users better utilize spot instances, but they are rarely used in reality. With a thorough analysis, we could uncover major hurdles when using the new datasets concerning the lack of historical information, query constraints, and limited query interfaces. To overcome them, we develop SpotLake, a spot instance data archive web service that provides historical information of various spot instance datasets. Novel heuristics to collect various datasets and a data serving architecture are presented. Through real-world spot instance availability experiments, we present the applicability of the proposed system. SpotLake is publicly available as a web service to speed up cloud system research to improve spot instance usage and availability while reducing cost.
\end{abstract}

\section{Introduction}
Cloud computing has changed the way in which we consume computing resources. One of the most significant changes is its novel on-demand billing model in which users pay for what they have used. Furthermore, excess computing resources are provided at a much lower cost than on-demand pricing, commonly referred to as a spot instance, which is provided by most public cloud vendors. Despite the lower price of spot instances, resources may be forced to be shut down as demand changes, and users should be prepared for a sudden interruption to running instances. Amazon Web Services (AWS), the leading cloud service provider, has been providing spot instance price history since the service inception to help users anticipate price changes and the possibility of instance interruption. The spot price change history dataset triggered extensive research~\cite{deconstructing-spot-instance, spot-analysis-javadi, stat-analysis-spot-price, spot-instance-analysis, spot-price-by-location} to optimize spot instance usage from various domains, such as big data processing~\cite{tr-spark}, deep learning~\cite{deepspotcloud}, and batch processing~\cite{spot-batch, spoton}. 

The spot price change history dataset is very popular in the literature for research and system implementation when using spot instances. However, in 2017, AWS changed its spot instance operation policy to make the price change less volatile~\cite{new-spot-price}. The change made most of the previous spot price analysis work obsolete and prevented the prediction of spot instance reliability using the spot price change history~\cite{spot-price-change-2017, spot-price-policy-change-2017-irwin}. Despite this, many spot instance-related works still rely solely on the price history data, which is irrelevant to the availability.

However, the spot price is not the only spot instance dataset, and public cloud vendors provide spot instance availability and interruption ratio datasets. For example, AWS provides the \emph{spot placement score} from $2021$, and it reflects the likelihood of spot request success, which can be an indication of timely availability. The spot instance advisor dataset\footnote{\url{https://aws.amazon.com/ec2/spot/instance-advisor}}, which was first released in $2015$ by AWS, provides the \emph{interruption ratio} of a spot instance during the preceding month as well as the cost saving ratio over on-demand price. Although these two most recent spot datasets provide useful information for spot instances, they have not received as much attention as the price dataset and have yet to be thoroughly evaluated.

Compared to the spot price dataset, whose historical and current information can be easily fetched programmatically, collecting the availability and interruption ratio datasets poses multiple challenges. Unlike the price dataset, the two new datasets do not include historical data. When creating a query, the spot placement score imposes numerous constraints. The interruption ratio dataset is accessible through the website, and it does not provide programmatic access. To expedite research in the related fields and enhance reliable spot instance usage by circumventing similar challenges that we have encountered during the data collection process, we have built a spot data archive service\footnote{\url{https://spotlake.ddps.cloud}} in which a user can access the historical dataset of spot instance availability, interruption frequency, cost savings, and spot price in a single place. 

Using the collected datasets, we conducted a thorough analysis to uncover the characteristics of various spot instance datasets. We discovered that distinct spot instance datasets present contradicting information quite often with low Pearson correlation coefficients~\cite{pearson-correlation-coefficient}, which might confuse spot instance users, and we showed which dataset is more credible through an empirical analysis. To show the applicability of the historical spot instance datasets, we built a simple machine learning model to predict spot instance interruption events, which were measured through real-world experiments, and even a simple model achieved better accuracy than considering only the current spot instance information. We are certain that further investigation and complex modeling from the research community using the diverse historical datasets can greatly improve usability while reducing interruptions. We believe the proposed spot dataset service can serve as a starting point to initiate new research to enhance spot instance reliability.

In summary, the major contributions are as follows.
\begin{itemize}
    \item First thorough composite analysis of spot instance availability, interruption ratio, and spot price dataset
    \item Real-world experiments to record the availability of spot instances, confirming the information validity provided by cloud vendors
    \item New insight, an abundance of spot availability datasets can result in more accurate prediction of spot reliability
    \item Provision of historical spot instance dataset and artifacts that are vital to initiating new research to enhance spot instance usage
\end{itemize}

\section{Transient Instances on Cloud and Datasets}
One of the key factors that led to cloud computing’s success was its elastic billing model. Before cloud computing, computing power had to be purchased in the unit of hardware. The on-demand pricing mechanism of cloud computing allows users to pay for resources only when they are needed. Aside from on-demand pricing, public cloud service vendors offer surplus computing resources at a much lower cost than on-demand instances. These are referred to as spot instances, and they include AWS Spot Instance, Microsoft Azure Spot Virtual Machines (VMs), and Google Cloud Spot VMs.

The spot cloud instance was first introduced by AWS in 2009. Since its inception, it adopts a market-driven auction through a uniform price and sealed bid mechanism. By the uniform price, all spot instance users pay for the same spot price regardless of a bid price, and users do not know the bidding price of other users (sealed bid)~\cite{deconstructing-spot-instance}. The service vendor determines the spot price based on supply and demand for cloud instances. The spot price varies depending on the instance type and availability zone. When an out-bid event occurs or idle resources become scarce, spot instances can be forced to shut down.

\begin{table}[t]
    \centering
    \begin{tabularx}{0.48\textwidth}{|c|X|}
    \hline
        \textbf{Status}   & \multicolumn{1}{c|}{\textbf{Description}}  \\ \hline
        Pending Evaluation & A valid spot request is submitted\\\hline
        Holding   &  Some request constraints cannot be met (price, location, resource availability, ...) \\\hline
        Fulfilled &  All the spot request constrains are met, and instance status being updated to running\\\hline
        Terminal & A spot request is disabled possibly by price outbid, resource unavailability, user, ...\\ \hline
    \end{tabularx}
    \caption{Possible spot instance request status and description}
    \label{table:spot-request-status}
\end{table}

The possible state of spot instance request is summarized in Table~\ref{table:spot-request-status}\footnote{\url{https://docs.aws.amazon.com/AWSEC2/latest/UserGuide/spot-request-status.html}}. Upon submission of a valid spot instance request, the request status becomes \emph{Pending Evaluation}. In the status, if any of the constraints cannot be met, the request status becomes \emph{Pending}. Possible reasons for the status include spot instance capacity not available in the requested available zone or bidding spot price too low. If all the constraints of a spot request are met, it becomes \emph{Fulfilled} status. In the status, an instance is started with configurations specified in a spot request. It might take a few minutes for an instance to start. A spot request with a running spot instance can become the \emph{Terminal} status for the following reasons: spot price out-bid or spot resource capacity not available, and they are generally referred to as spot instance interruption. A user can also terminate an instance voluntarily.

Owing to the lack of spot instance reliability, users should prepare a plan to deal with an instance interruption. To help users better utilize spot instances, service vendors provide diverse information, such as spot instance price, current spot instance availability, and statistics of spot instance interruption ratio over the prior period.

\subsection{Spot Instance Price}\label{sec:spot-price-history}
AWS provides spot instance price change history dataset through its website and Command Line Interface (CLI) library to allow programmatic access. Users can specify the start and end times of queries as well as availability zones and instance types. The returned output includes the timestamp at which a spot price changes as well as the changed spot price at the time. When using spot instances, the history of the spot price for the previous three months provides insightful information, and many studies have been conducted using the dataset. Statistical analysis of the spot price change helps users to better understand the spot instance market~\cite{draft-spot-instance-guarantee-from-spot-price, deconstructing-spot-instance, spot-analysis-javadi, stat-analysis-spot-price, spot-instance-analysis, spot-price-by-location, spot-instance-for-hpc}. Using the spot price change history, many studies have been conducted to propose an optimal bidding algorithm~\cite{how-to-bid-cloud, how-not-to-bid-cloud, spot-bidding-infocom, spot-low-bidding-hpdc, lstm-price-predict, optimal-spot-bidding-cloud, spot-bidding-with-deadline-cloud, minimum-bidding-estimation-ucc, spot-price-prediction}. 

The active and wide usage of spot price change data was feasible because the data update frequency was timely and users could estimate the likelihood of instance interruption and cost savings in AWS. However, in 2017, AWS changed its spot instance operation policy~\cite{new-spot-price}. In the new policy, the spot price changes less frequently and is thus more stable~\cite{spot-price-policy-change-2017-irwin, spot-price-change-2017}. However, spot prices in the new policy no longer represent interruption events. Before the update, a user could predict the interruption events by comparing the bid and spot prices. Now, following the update, the spot price does not accurately reflect the surplus of idle computing resources, particularly when it is low. As a result, when an advertised spot price remains lower than the bid price, the instance interruptions can still occur, making the results of previous related research work obsolete.

\subsection{Spot Instance Interruption Ratio}\label{sec:spot-instance-advisor}
Instance interruptions are a major concern for most spot instance users. To help users estimate the probability of interruption, cloud service vendors provide the interruption ratio. AWS's Spot Instance Advisor provides the rate at which spot instances have been interrupted in the preceding month. The interruption frequency data are divided into five categories: less than $5\%$, between $5\%$ and $10\%$, between $10\%$ and $15\%$, between $15\%$ and $20\%$, and more than $20\%$. The interruption ratio data are provided per a combination of instance type and region, which is a coarser-grained manner than that of spot price. The service is officially accessible via the website only, and it does not support the programmatic access.

\subsection{Spot Instance Availability}\label{sec:spot-placement-score}
The spot instance interruption ratio dataset reflects the reliability during the preceding month, and it might not reflect the current spot resource availability in a timely manner, which can directly impact the success of a spot instance request. To provide timely spot instance availability, AWS offers the Spot Placement Score service~\cite{spot-placement-score-start}. The placement score's primary goal is to assist users in estimating the likelihood of a successful spot request before launching an instance type in a specific availability zone. The internal details of how the metric is calculated are not publicly available. Externally, it takes the desired instance types, regions, and target capacity as arguments and returns a placement score ranging from $1$ to $10$, with a higher score indicating a greater likelihood of spot request success. The spot placement score can be accessed via the website and CLI.

\section{Spot Instance Datasets Collection}
Unlike the spot instance price dataset, which provides historical information via CLI, the other two recent spot instance data sources presented in Sections~\ref{sec:spot-instance-advisor} and \ref{sec:spot-placement-score} have a few shortcomings in the data itself and access medium with query constraints that must be overcome. 

\subsection{Challenges in Data Collection}
\textbf{No historical data}: For both the spot instance interruption ratio and availability datasets, users can query only the current value, and no historical information is provided. Meanwhile, up to three months of spot price history data are provided, and long-term data history was a valuable source for optimizing spot instance usage~\cite{spot-analysis-javadi, spot-price-prediction}. Thus, to allow the spot instance interruption ratio and availability datasets to enhance spot instance usage, historical data should be provided allowing further analysis from academia and industry.

\textbf{Query limitation}: The spot instance availability information expressed by the spot placement score imposes various limitations in the query, and they are described in its official document page~\cite{splot-placement-score-limitation}. The most significant limitation is the number of unique queries allowed in $24$ hours. This can be a serious drawback, especially when a user wants to investigate the availability of multiple instance types across several regions. According to our empirical analysis, an account can issue a maximum of $50$ unique queries in $24$ hours. The uniqueness of a query is determined by the combination of regions, instance types, and the number of desired instances. An aggregated placement score value for the specified instance types in a query is returned for each region in the argument for each unique query. Issuing the same query multiple times does not count against the query limit. Thus, one can issue $50$ unique queries multiple times to record the values changes.  

The spot placement score differs for distinct availability zones even if they are located in the same region. In a query, a user can specify an option of \emph{SingleAvailabilityZone} as \emph{true} to get score values for each availability zone in a region specified in the query. Another query regulation is imposed for the number of returned placement scores, which is limited to $10$. For example, if a query specifies multiple regions with the \emph{SingleAvailabilityZone} option as \emph{true}, there can be more than $10$ placement scores for different availability zones. In such a case, only $10$ placement scores with larger scores are returned.

At the time of this writing, there are about $547$ instance types, $17$ regions, and $63$ availability zones in AWS. To scan the spot placement scores for all possible instance type and region combinations, $547 \times 17 = 9,299$ queries should be executed at most. Given the limit of $50$ unique queries per account, it is impossible to obtain scores for all instance type region combinations. Users can optimize a query by specifying multiple regions in a query, but the number of placement score results returned from a query is still limited to $10$. A large number of possible spot placement score query dimensions regarding instance types, regions, availability zones, and the number of instances necessitates the use of a specialized data service to provide comprehensive, timely information.

\textbf{Limited query interface}: Among the many ways to use and operate cloud resources, programmatic access is preferred over a management console that uses a graphical user interface~\cite{cloud-cli}. From the context, spot price history and spot placement score data are provided through CLI, which allows programmatic management. However, the interruption ratio information is natively supported only from the management console and hurts operability~\cite{designing-data-intensive-applications} without programmatic access.

\subsection{Data Collection Methodology}\label{sec:data-gathering-heuristic}
To improve efficiency when using spot instances, it is important to broaden target instance types if a workload does not have a specific hardware requirement. Furthermore, if a workload does not have a specific geographical requirement, the possibility of cost saving can be further improved~\cite{deepspotcloud,spot-price-by-location}, and building a compute cluster with heterogeneous spot instances is proven to be an efficient solution for data-parallel analysis tasks~\cite{heterogenoue-spot-instance-application}. Timely and detailed spot instance status information is required to build a cost-optimal cloud environment while broadening candidate spot resources.

To optimally query the spot instance availability dataset, we must organize an instance type with multiple regions, each of which may have a different number of availability zones. We must also consider that not all availability zones in a region support a specific instance type. To run spot placement queries as efficiently as possible, we created a nested dictionary whose key is an instance type and the corresponding value is another dictionary whose key is the region and the value is the number of availability zones that support the instance type in the outer dictionary's key. The problem can be simplified to a bin-packing algorithm~\cite{bin-packing-algorithm}, which collocates regions together while making the sum of the number of availability zones of each region the maximum number of returned results. After the problem abstraction, we used Google OR-Tools~\cite{ortools} to solve the bin-packing problem. Among possible libraries, we used COIN-OR Branch-and-Cut (CBC)~\cite{coin-or-branch-cut} implementation, which is a mixed-integer programming solver. After packing multiple regions with a single instance type to be requested in a query, we decreased the total number of required queries from $9,299$ to $2,226$ which is about a $4.5 \times$ improvement.

\begin{figure}
    \centering
    \includegraphics[width=0.47\textwidth]{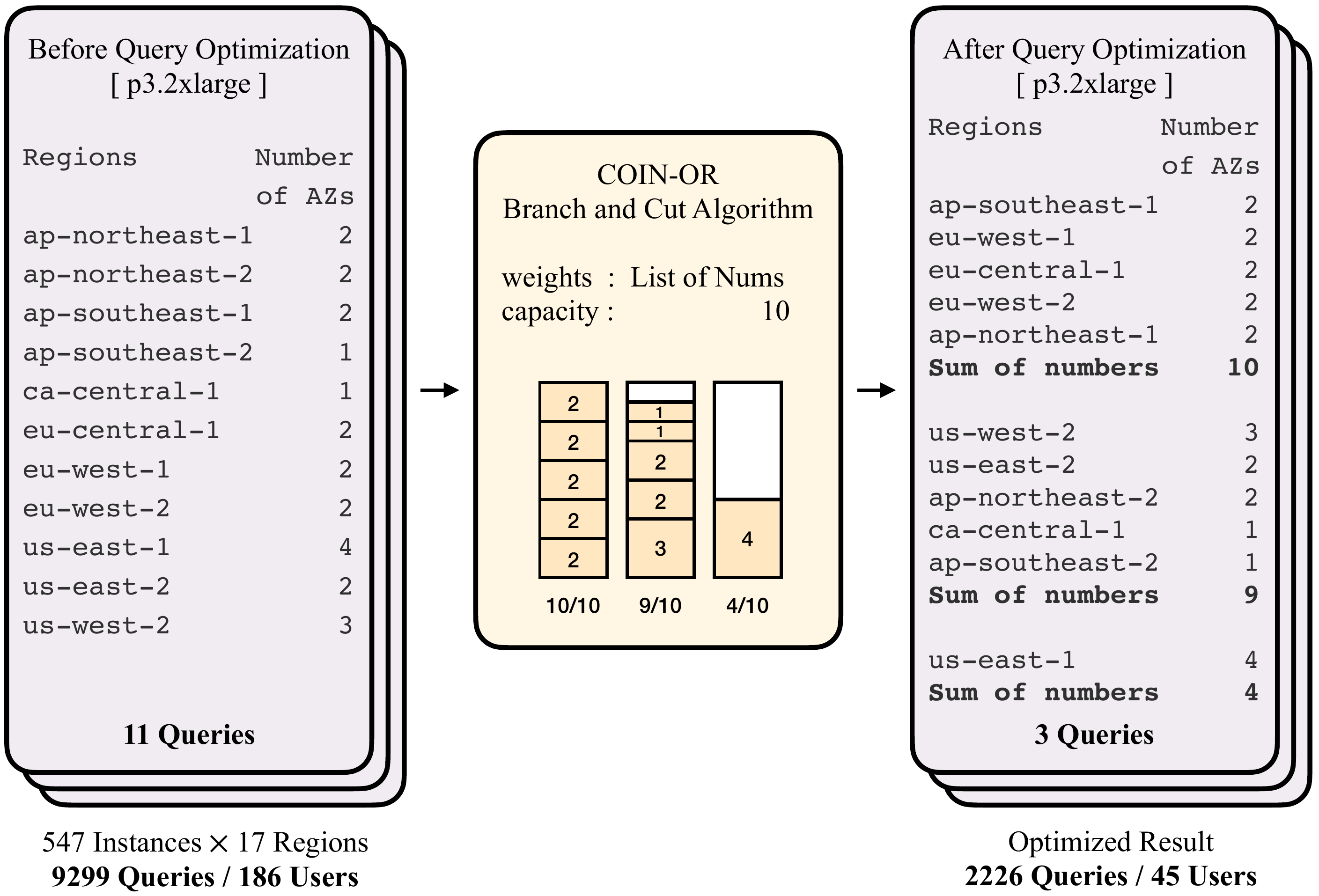}
    \caption{Spot placement score query optimization using bin-packing algorithm (mixed integer programming solver)}
    \label{fig:query-opt-architecture}
\end{figure}

Figure~\ref{fig:query-opt-architecture} explains the spot placement query optimization example with a sample instance type of \emph{p3.2xlarge}. The regions and number of availability zones that support the instance type are shown on the left side of the figure. Following the bin-packing execution, multiple regions are grouped into a single query to reduce the total number of queries.

\section{Implementation of Spot Data Archive Service}
\begin{figure}
    \centering
    \includegraphics[width=0.44\textwidth]{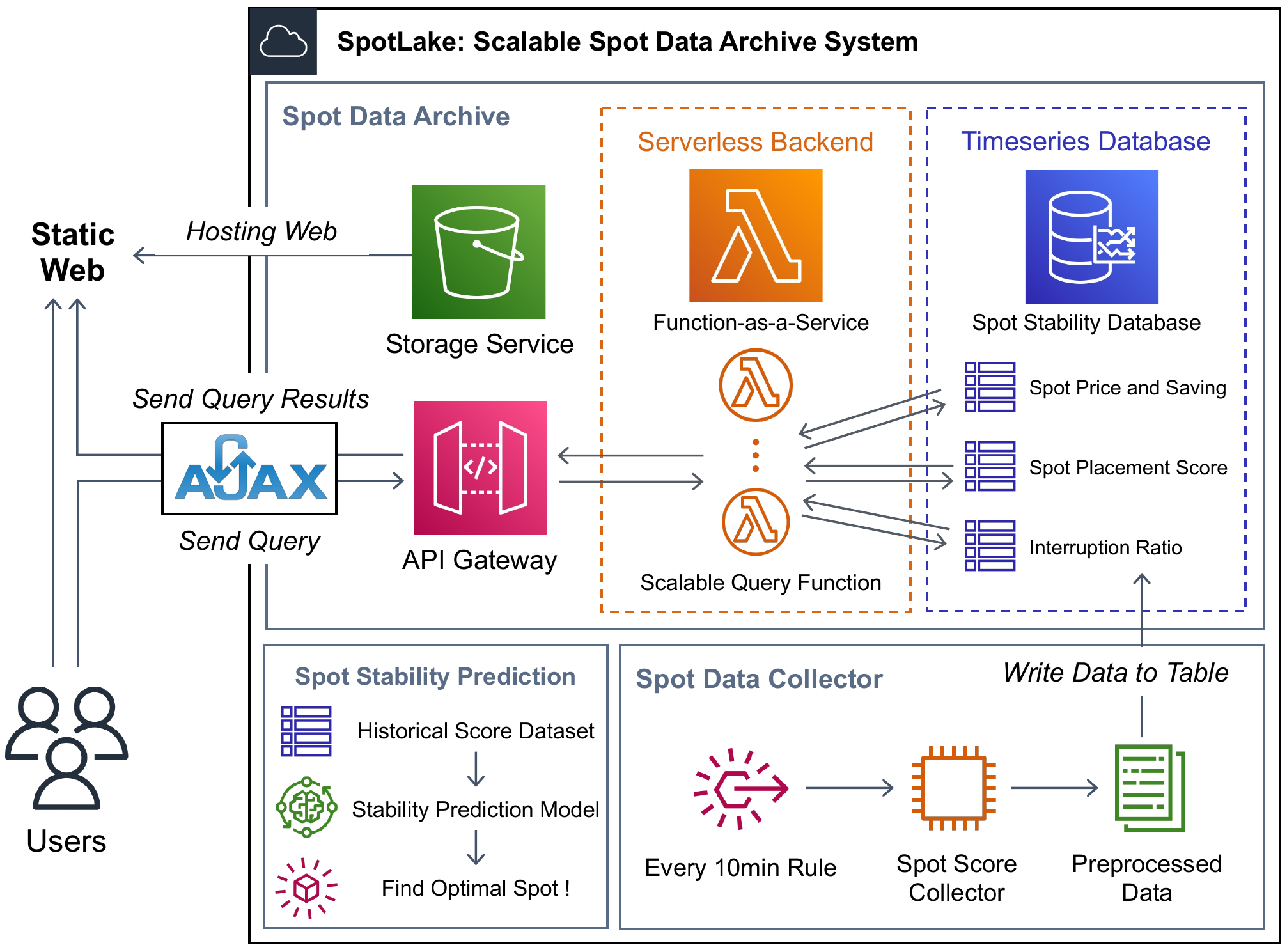}
    \caption{Data archive service implementation adopting a serverless architecture}
    \label{fig:spot-service-architecture}
\end{figure}

Challenges in collecting diverse historical spot instance datasets for heterogeneous instances located globally might slow down new study outcomes from the cloud system research community. To overcome the shortcomings, we implement a data archive web service\footnote{\url{https://spotlake.ddps.cloud}} that provides historical information of the spot instance availability, the interruption ratio, and the cost savings over on-demand instances. The architecture of the implemented web service is shown in Figure~\ref{fig:spot-service-architecture}. To relieve resource management burdens of the application servers, we adopted a serverless architecture where applicable. The spot data collector server periodically executes collection tasks for different data sources. The spot advisor dataset does not support programmatic access, and we used the SpotInfo~\cite{spotinfo} tool to collect the dataset programmatically. The spot instance advisor dataset can be queried with a single execution, whereas the spot placement score necessitates multiple queries with various entities. The spot dataset can be well represented using a time-series format, and we use an Amazon \emph{Timestream} database, a fully managed time-series database service.

The front-end service is also implemented by adopting a serverless architecture where static files are served from an object storage service (Amazon \emph{S3}), and the dynamic contents in a webpage are updated in real-time using an AJAX protocol. A user's request with various query parameters is delivered via \emph{API Gateway} and passed to \emph{Lambda}, which fetches necessary datasets from the \emph{Timestream} database. The current implementation keeps a historical dataset of the spot placement score and instance advisor dataset for all regions and instance types. Users can query specifying the timestamp, regions, availability zones, and instance types.

\section{Spot Instance Data Analysis}\label{sec:analysis}
Many of the prior works in the literature have analyzed the spot price history data thoroughly. However, to the best of the authors' knowledge, no prior work has analyzed the spot instance availability and interruption ratio datasets. As the practicality of the spot price dataset weakens~\cite{spot-price-policy-change-2017-irwin}, we are confident that a thorough analysis of other available spot instance information is critical to better utilize spot resources.

For the analysis, we used the current spot instance availability dataset provided by the spot placement score and the interruption ratio dataset from January 1, 2022 to June 30, 2022, for a total of $181$ days. The data were collected every $10$ minutes. In the analysis, we used the term spot placement score to represent the current spot instance availability. The interruption ratio, which was included in the spot instance advisor dataset, was provided as a categorical value from the lowest frequency of less than $5\%$ to the highest frequency of more than $20\%$. To improve our analysis, we converted the categorical value to a score value by matching the range with that of the spot placement score, which is between $1.0$ and $3.0$. We set the lowest interruption frequency to $3.0$ and the highest interruption frequency to $1.0$ in the score representation. There are three more categorical values in between, and we assign them $2.5, 2.0, $, and $1.5$, ranging from lower to higher interruption frequency. The converted interruption frequency information is referred to as the interruption-free score. A higher interruption-free score, like a higher spot placement score, indicates better spot instance availability.

\subsection{Spot Placement and Interruption-free Score}
\begin{figure}[t]
    \centering
    \subfloat[Spot placement score]{
        \includegraphics[width=0.45\textwidth]{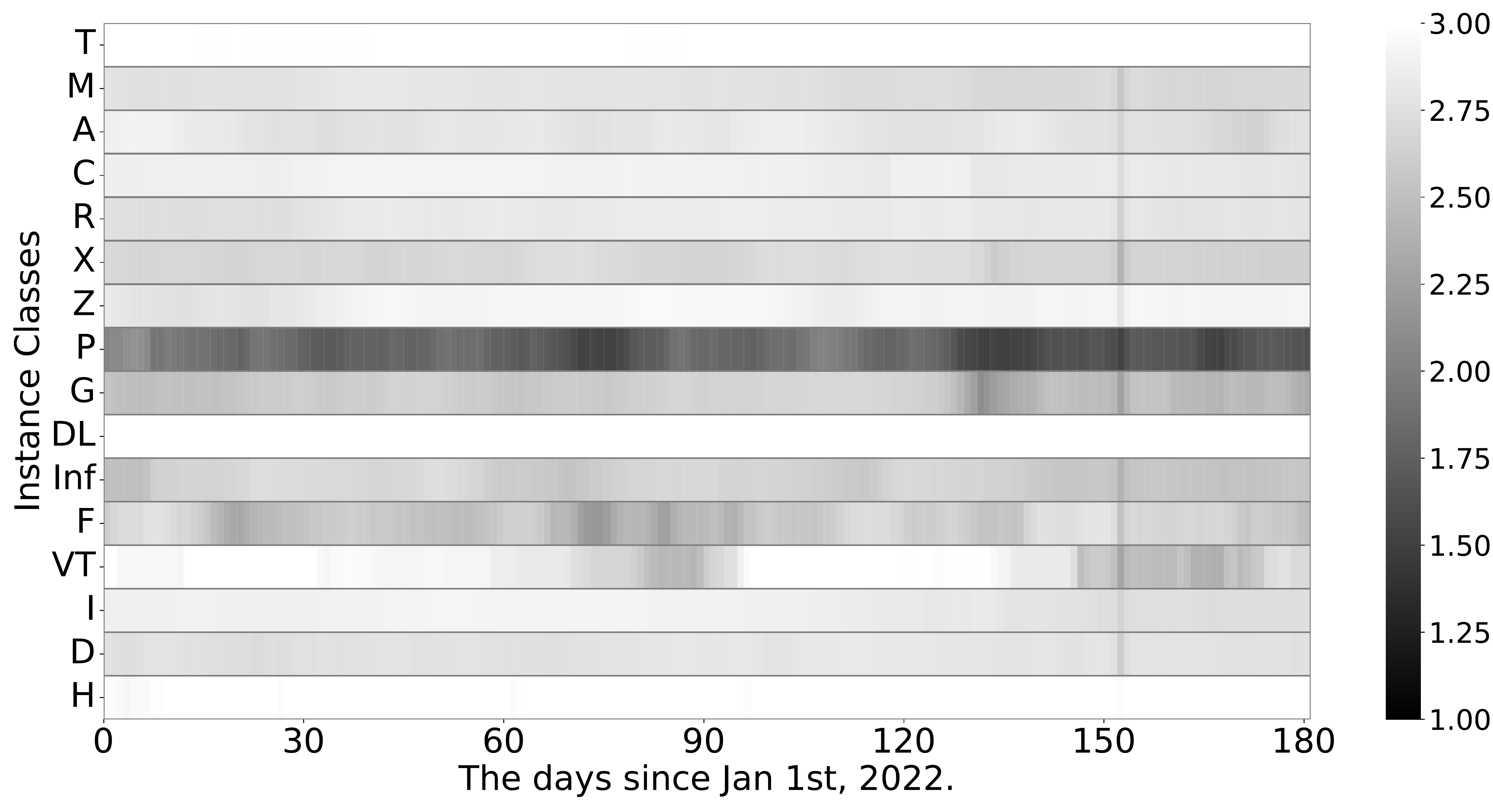}
        \label{fig:sps-by-instance-types}
    }\\
    \subfloat[Interruption-free score]{
        \includegraphics[width=0.45\textwidth]{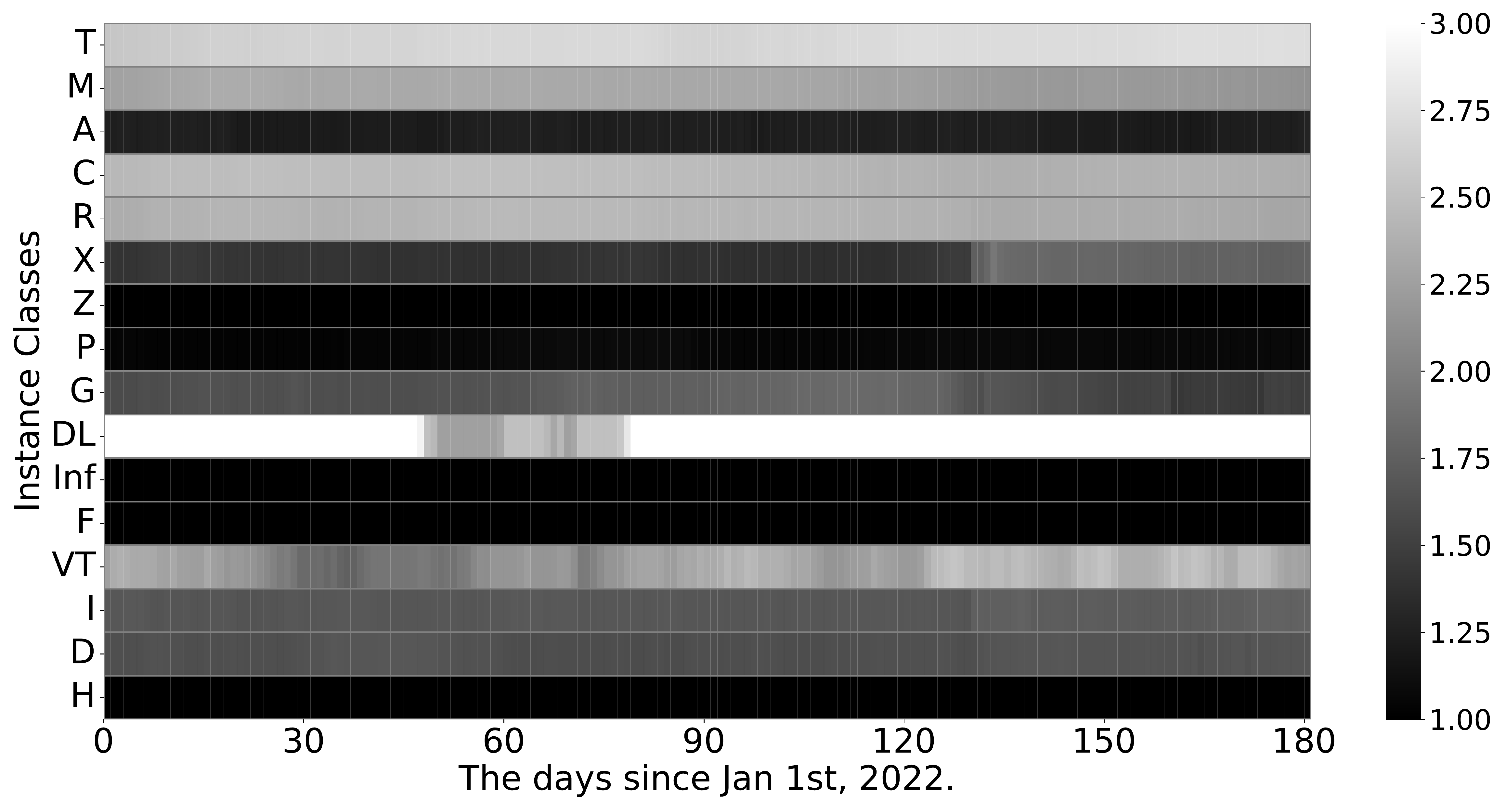}
        \label{fig:if-by-instance-types}
    }
    \caption{Temporal variations of spot instance scores}
    \label{fig:scores-by-instance-types}
\end{figure}

Figure~\ref{fig:scores-by-instance-types} presents temporal changes of spot placement score (Figure~\ref{fig:sps-by-instance-types}) and interruption-free score (Figure~\ref{fig:if-by-instance-types}) using a grayscale heatmap. The brighter colors express higher spot placement and interruption-free scores, which imply higher spot instance availability. The horizontal axis shows the elapsed days since the data collection start date. The vertical axis represents instance classes offered by AWS. They are shown in the order of \emph{general} instance family (T, M, A), \emph{compute-optimized} instance family (C), \emph{memory-optimized} instance family (R, X, Z), \emph{accelerated-computing} instance family (P, G, DL, Inf, F, VT), followed by \emph{storage-optimized} instance family (I, D, H). In each instance class, daily average score values are calculated.

The figures show that the spot placement score shows a much lighter color (higher score) than the interruption-free score. Overall, the average spot placement score across all the instance types is $2.8$, and that of the interruption-free score is $2.22$. Among many instance types, the \emph{accelerated-computing} instance family has the lowest spot placement and interruption-free scores, which are $12.07\%$ and $34.98\%$ lower than the average scores, respectively. We can infer that such characteristics stem from the recent popularity of the deep learning in which specialized hardware is widely used both for training and inference tasks~\cite{dnn-workload-sensetime, dnn-workload-character-microsoft}. Among the \emph{accelerated-computing} instance types, the \emph{DL} instance shows high spot placement and interruption-free scores. The instance type provides a Gaudi processor, which is special-purpose hardware for Deep Neural Net (DNN) training and inference developed by Habana Labs~\cite{gaudi-accelerator}. It was released recently, and we assume that the eco-system for DNN development using the instance type is not yet mature, which may explain the low usage so far. For instance types with GPU devices, the \emph{G} instance class shows higher scores than \emph{P} instance types. The \emph{G} instance type equips NVIDIA T4 GPU (\emph{G4dn}) or AMD V520 (\emph{G4ad}), and the \emph{P} instance type equips NVIDIA Tesla V100 GPU (\emph{P3}) or Tesla A100 GPU (\emph{P4}). Comparing the two instance types, the \emph{P} instance type shows better performance for DNN tasks~\cite{aws-ec2-g-p-comparison}, but the \emph{G} instance type is more affordable with a lower hourly price. We presume that the better affordability and cost-performance ratio of \emph{G} instance types result in higher resource capacity and increased possibility of surplus resources. Concerning the temporal score changes, neither spot placement nor interruption-free scores show significant score variations as the date changes on the horizontal axis. For the spot placement score, a sudden decrease was uncovered around June 2, 2022. Further investigation revealed score adjustments for most instance types during the time, which might have resulted from the spike in the spot instance usage.

\begin{figure}[t]
    \centering
    \subfloat[Spot placement score]{
        \includegraphics[width=0.45\textwidth]{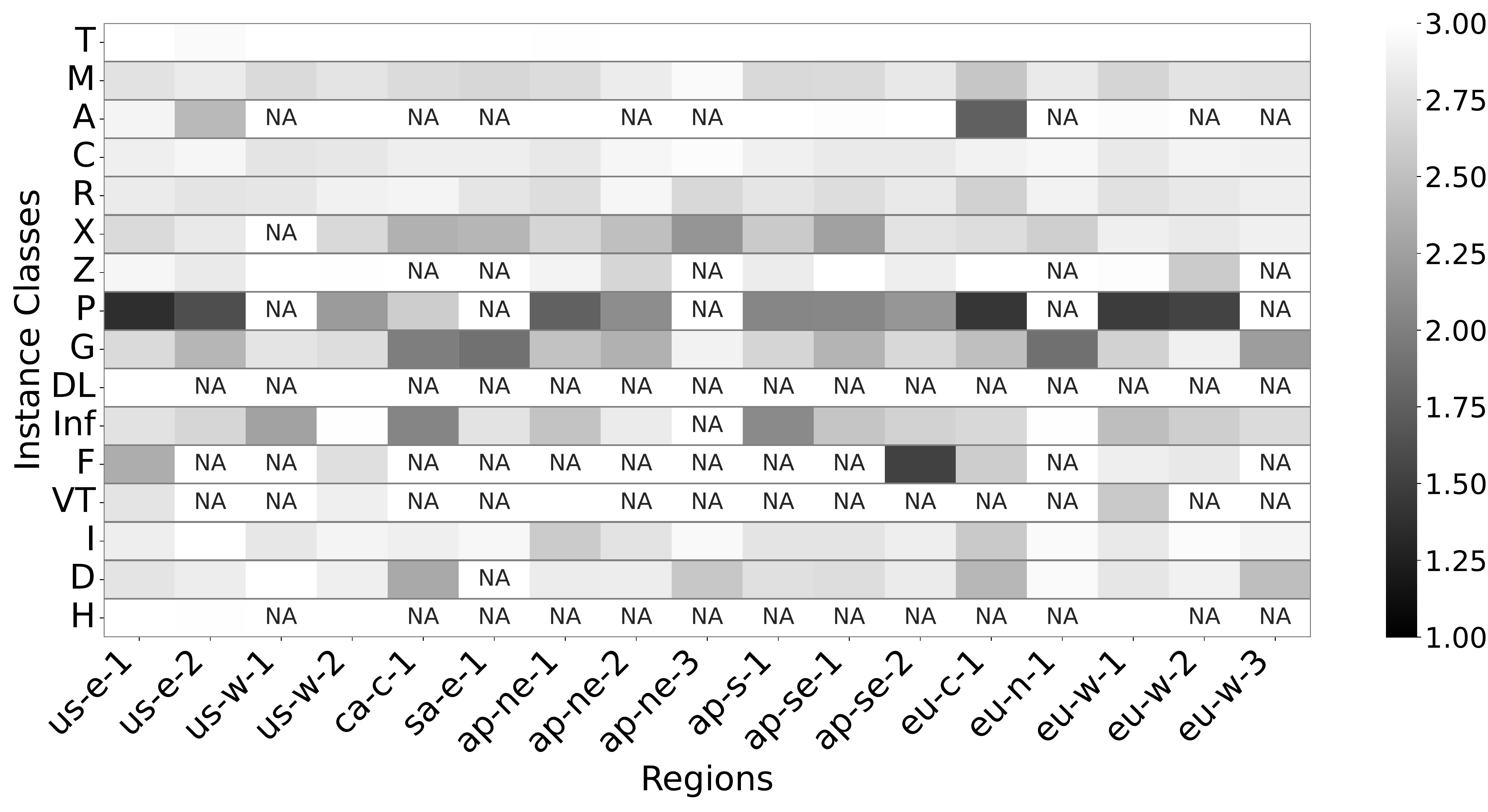}
        \label{fig:sps-by-regions}
    }\\
    \subfloat[Interruption-free score]{
        \includegraphics[width=0.45\textwidth]{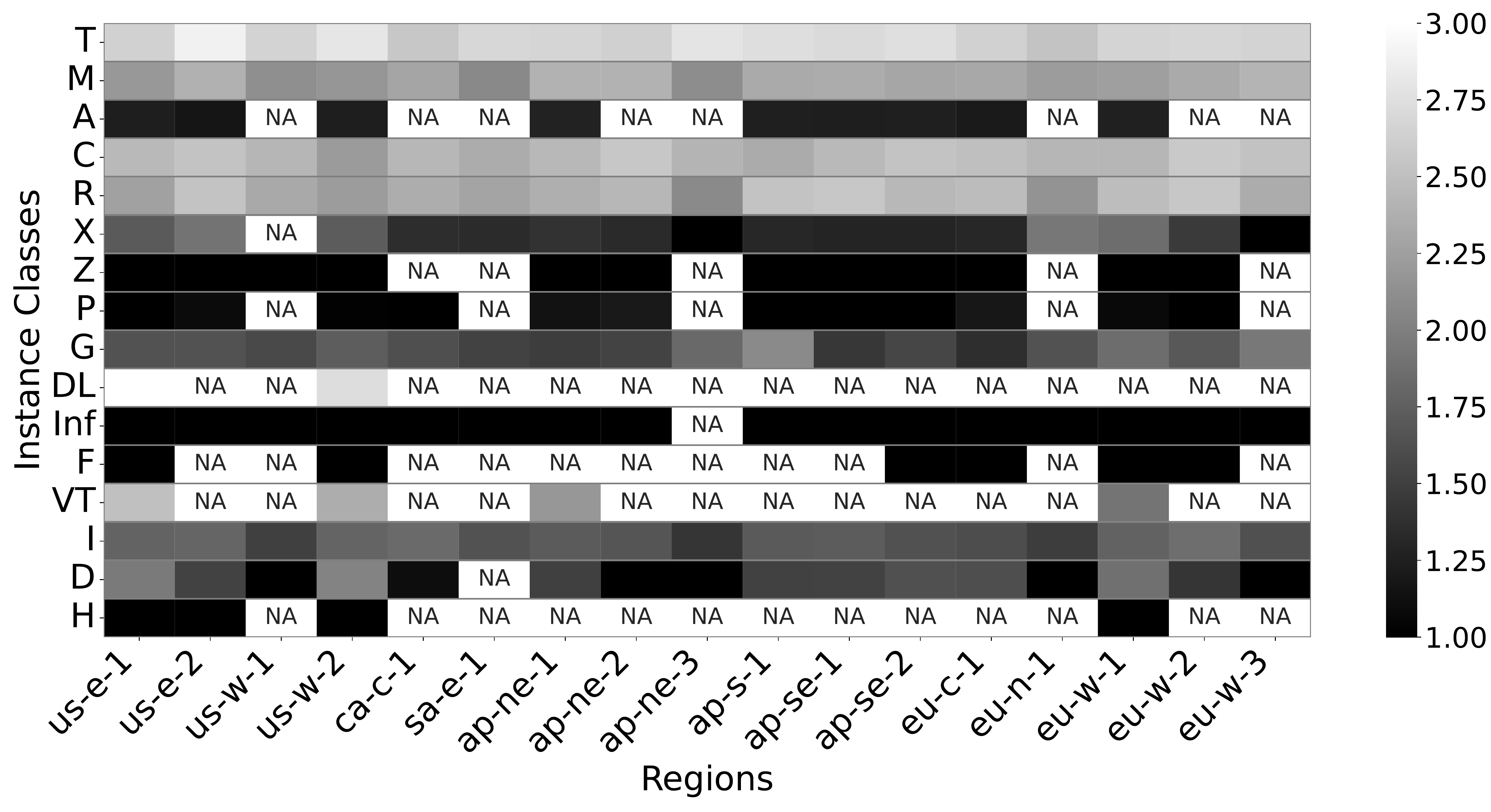}
        \label{fig:if-by-regions}
    }
    \caption{Spatial variation of spot instance scores}
    \label{fig:scores-by-regions-instance-types}
\end{figure}

To observe spatial characteristics of the spot placement score for different instance classes, Figure~\ref{fig:scores-by-regions-instance-types} shows the difference in scores between regions depending on spot placement score (Figure~\ref{fig:sps-by-regions}) and interruption-free score (Figure~\ref{fig:if-by-regions}) with a grayscale heatmap. To cover a wide range of instance types, we chose $17$ regions with the greatest number of instance types supported, which are displayed on the horizontal axis. A region code is expressed in the \emph{continent-coordinate-id} combination. The color scale and the instance classes presented on the vertical axis are the same as those in Figure~\ref{fig:scores-by-instance-types}. Instance types that are not supported in a specific region are marked as \emph{NA}. From the figures, we can visually observe a higher degree of score variations across different regions, which coincides with the conclusions from previous work~\cite{spot-price-by-location}. Among \emph{accelerated-computing} instance classes, the general-purpose GPU devices (\emph{G} and \emph{P}) show relatively lower scores for most regions. Thus, if users build a DNN environment for a specific purpose, either training or inference, they can use a more reliable spot instance environment by using special-purpose instance types located globally~\cite{deepspotcloud}, such as \emph{DL} for training and \emph{Inf} for inference.

To understand the score distribution of the spot placement and interruption-free scores, Table~\ref{table:sps-if-value-distribution} shows the percentage of values observed during the measurement period. According to the table, most spot placement scores are $3.0$, and only $8.31\%$ of spot placement scores are $1.0$, indicating a high likelihood of the spot request being successful. The interruption-free score, as opposed to the spot placement score, exhibits a more uniform distribution across distinct values. About $33.05\%$ of cases show less than $5\%$ interruption frequency (score value : $3.0$), but $20.84\%$ of cases show an interruption ratio of more than $20\%$ (score value : $1.0$). From the scores' value distribution, we can roughly identify the difference between the two spot datasets.

\begin{table}[t]
    \centering
    \begin{tabularx}{0.45\textwidth}{|c||X|X|}
    \hline
            \makecell{Value} & \makecell{Spot placement score} & \makecell{Interruption-free score}  \\ \hline
            
        3.0 & \makecell{87.88\%}    & \makecell{33.05\%} \\
        2.5 & \makecell{NA}         & \makecell{25.92\%} \\
        2.0 & \makecell{3.81\%}     & \makecell{13.86\%} \\
        1.5 & \makecell{NA}         & \makecell{6.33\%} \\
        1.0 & \makecell{8.31\%}     & \makecell{20.84\%} \\ \hline
    \end{tabularx}
    \caption{Value distribution of spot placement score and interruption-free scores (higher value implies more stability)}
    \label{table:sps-if-value-distribution}
\end{table}

\begin{figure}
    \centering
    \includegraphics[width=0.45\textwidth]{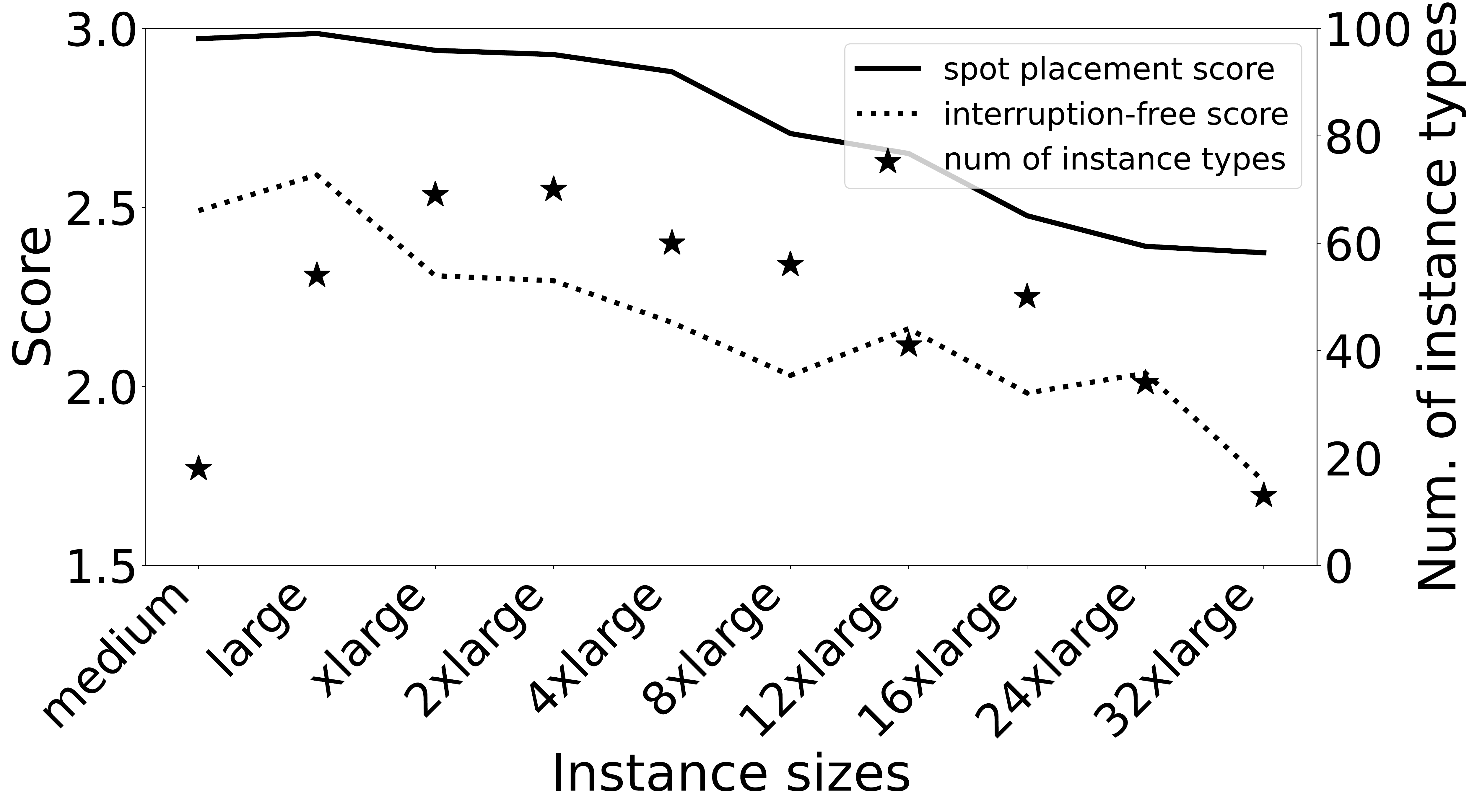}
    \caption{The spot placement and interruption-free scores grouped by the instance sizes}
    \label{fig:scores-by-instance-sizes}
\end{figure}

Figure~\ref{fig:scores-by-instance-sizes} shows the spot placement and interruption-free scores for different instance sizes, which are expressed on the horizontal axis. The solid line represents the spot placement score, and the dotted line represents the interruption-free score, both of which are represented by values on the primary vertical axis. For the instance sizes expressed on the horizontal axis, we choose ones with greater than $10$ corresponding instance types. We discover that for instance sizes with a low number of instance types, the average score value is determined by only a few instance types and does not adequately reflect the impact of sizes. The number of instance types is expressed using a star marker whose value is shown on the secondary vertical axis. From Figure~\ref{fig:scores-by-instance-sizes}, we can observe that as the instance size increases, both the spot placement and interruption-free scores decrease, which coincides with Kadupitige et al.~\cite{google-cloud-empirical-preemption}. Larger instance sizes necessitate more computing resources and are more likely to cause resource fragmentation without instance migration~\cite{vm-placement-fragmentation}. The lower flexibility of the larger size can result in a low possibility of surplus resources and lower availability scores.

\textbf{Key findings}: For spot placement and interruption-free scores, the spatial diversity is more noticeable than the temporal diversity, and it is recommended to distribute spot instance usage across different regions~\cite{deepspotcloud}. The \emph{accelerated-computing} instance family shows noticeably lower availability than other instance families. Requesting a smaller size instance type is more likely to succeed with a lower number of interruptions.

\subsection{Spot Placement Score With Diverse Parameters}
When querying spot placement scores, users can specify multiple instance types for a composite spot placement score. To understand the characteristics of composite instance type queries, we compared the placement score of a query that specifies multiple instance types and scores of multiple queries, each specifying a single instance type. The goal of this analysis was to determine how a single instance spot placement score affects the score when the instance types are queried together. According to the official spot placement score document, the score value ranges from $1$ to $10$. However, the maximum returned score in our experimental queries that specified only a single instance type was $3$. From the observation, we assumed that the maximum spot placement score, $10$, would be returned when a query specifies multiple instance types, and the placement score of multiple instance types might be the sum of the individual instance's scores. 

\begin{figure}
    \centering
    \includegraphics[width=0.4\textwidth]{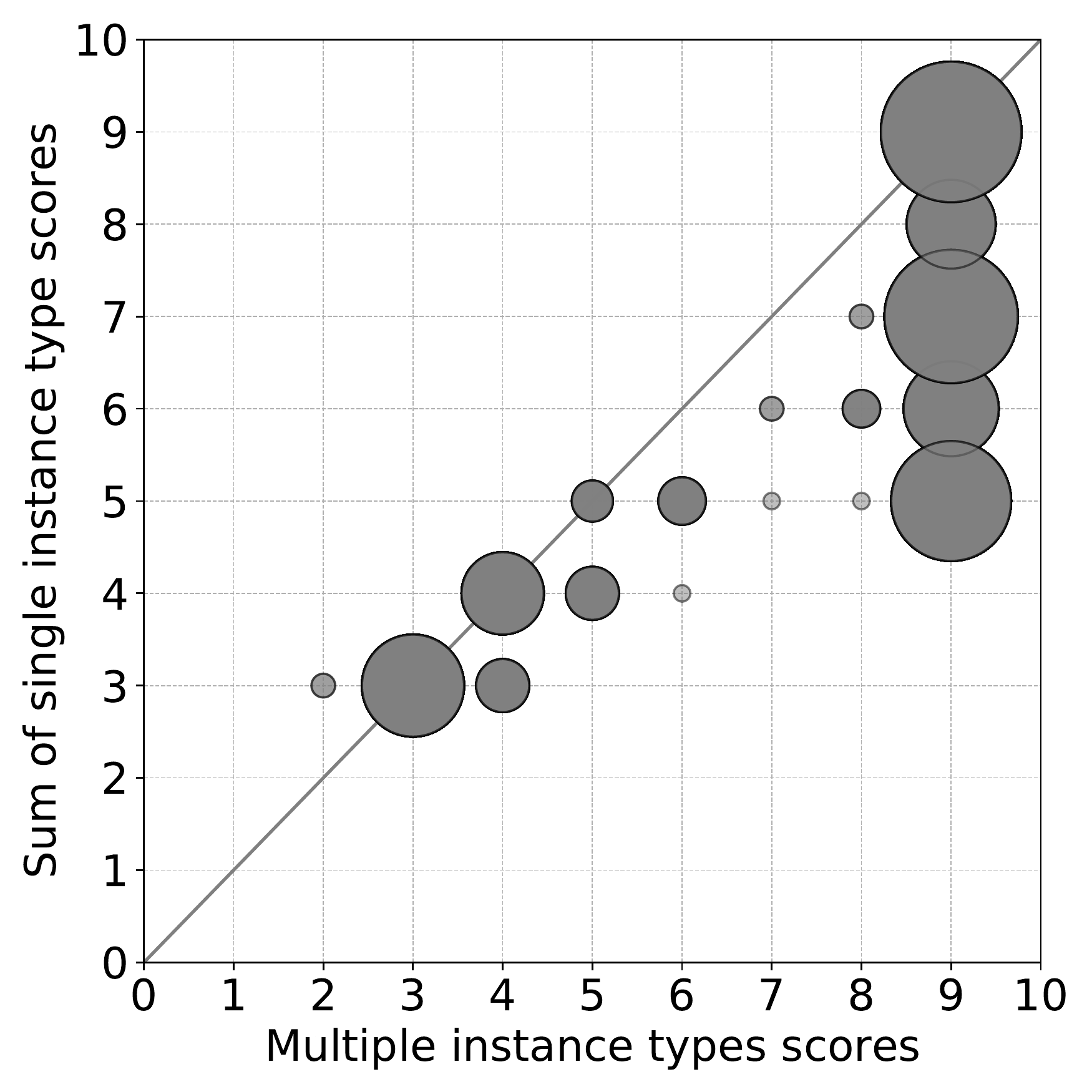}
    \caption{Spot placement score composite instance type query}
    \label{fig:sps-query-instance-type-composition}
\end{figure}

To check the validity of the hypothesis, we issued multiple queries that specified three arbitrary instance types. Figure~\ref{fig:sps-query-instance-type-composition} presents the returned placement score of a query with multiple instance types on the horizontal axis. The vertical axis presents the summed spot placement scores when a query is made separately for each instance type. To uniformly distribute the sum of individual spot placement scores, we chose the same number of instance type and availability zone combinations in each summed score value, which ranged from $3$ (all the three instance scores were $1$) to $9$ (all the three instance scores were $3$). The figure is presented in scatter plot format whose radius represents the frequency of occurrences. 

In the figure, we add a line of $y=x$ with a slope of $1$ to indicate a case in which the returned spot placement score of multiple instance types is the same as the sum of the individual spot placement scores of each instance type. In the experiments, about $38.81\%$ of cases are this type of case. The circles to the lower-right of the $y=x$ line indicate cases in which the spot placement score of multiple instance types is larger than the sum of individual scores, and about $60.62\%$ of cases are this type of case. Based on the results of this experiment, we can conclude that the sum of the individual scores can be the smallest of the composite instance type spot requests. We observed two cases in the experiments in which the composite instance type score was less than the sum of the individual scores, which we considered to be exceptions.

\begin{figure}
    \centering
    \includegraphics[width=0.45\textwidth]{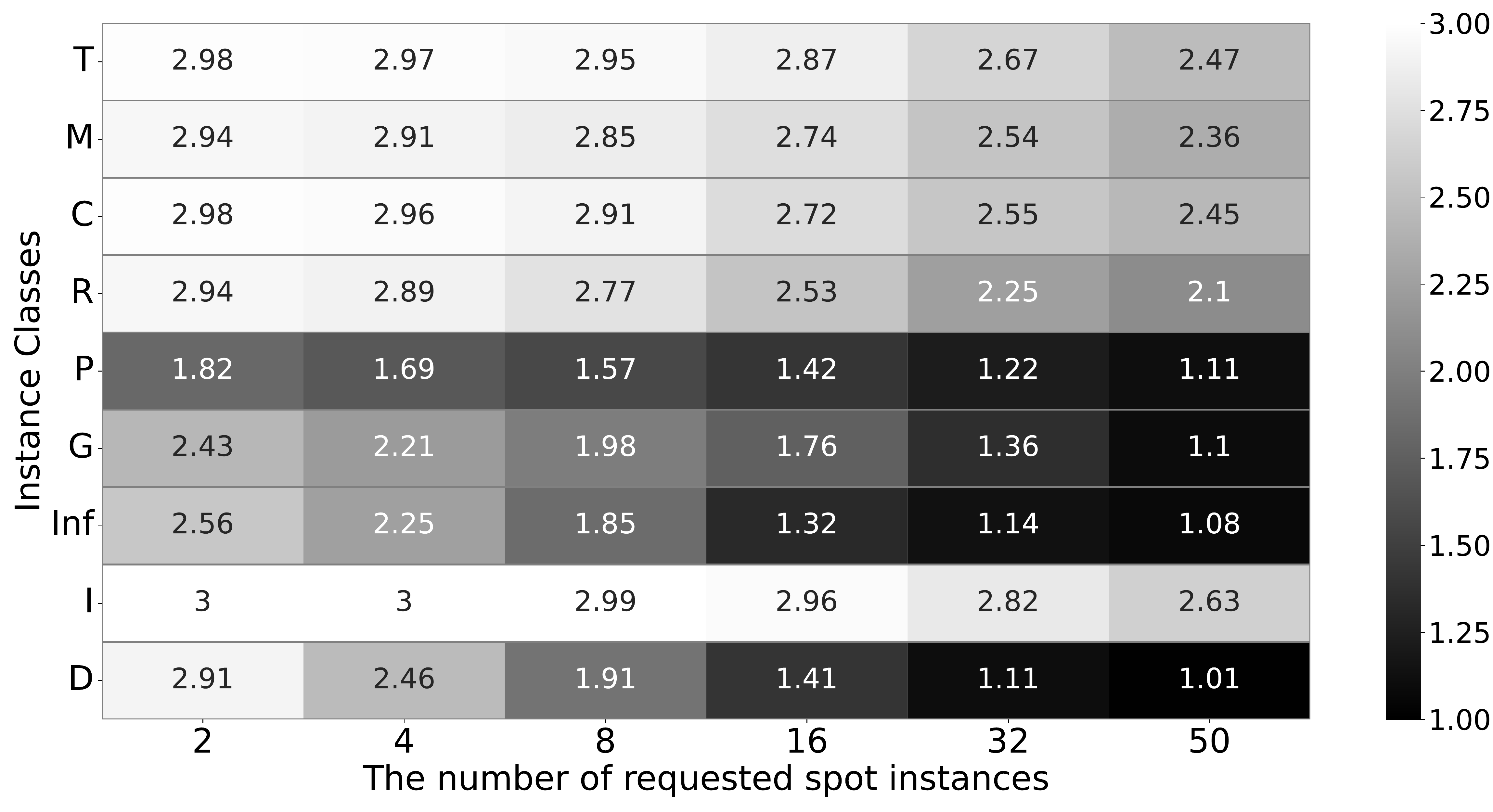}
    \caption{Changes of spot placement scores regarding different number of requested resources}
    \label{fig:sps-changes-capacity-change}
\end{figure}

In a spot placement score query, users can specify the number of spot instances to request, and Figure~\ref{fig:sps-changes-capacity-change} presents how the spot placement score changes when a large number of instances is requested. The number of requested instances in a query is shown on the horizontal axis, and the instance classes are shown on the vertical axis. In the experiments, we selected a few representative instance types in each instance family. To control the total number of queries, we used only the \emph{xlarge} size where applicable. For instance, for types that did not have the size, \emph{P4}, we used the smallest possible size. 

Intuitively, specifying more instances in a spot request lowers the chances of the fulfillment, and we can discovered such a pattern. The ratio of spot placement score decrease differed quite significantly across distinct instance types. For example, instance types in the \emph{accelerated-computing} family, \emph{P, G} and \emph{Inf}, showed significant score drops when a large number of resources was requested. The \emph{D} instance type, which belongs to the \emph{storage-optimized} family, also presented a noticeable score drop. Such instance types are armed with specialized hardware internally in a host, such as GPU devices in \emph{P} and \emph{G}, AWS Inferentia chips in \emph{Inf}, and large local storage disks in \emph{D} instances, and the supply of such resources might be lower than that of other general-purpose instance types.

\textbf{Key findings}: The sum of the individual spot placement scores of different instances can be regarded as the minimum score of the composite spot placement scores. Requesting a large number of spot instances in \emph{accelerated-computing} shows larger availability drops than that in general instance families. 

\subsection{Correlations Among Multiple Spot Datasets}
So far, we have analyzed the spot placement score and interruption-free score independently. The two datasets are generated in real-time, and the spot instance price dataset is also available at the same time. From the spot users' perspective, three spot data sources exist, and it can be challenging to decide which dataset provides the most accurate information to infer the spot instance availability, especially when distinct datasets imply contrasting information. To understand the correlation among the spot instance price, spot placement score, and interruption-free score, we use the Pearson correlation coefficient~\cite{pearson-correlation-coefficient} of the two datasets' combination among the spot placement score, interruption-free score, and spot price. The Pearson correlation coefficient of two variables, $X$ and $Y$, is calculated as follows.
\begin{equation*}
    R_{XY} = \frac{\sum_{t}^{T}(X_t-\overline{X})(Y_t-\overline{Y})}{\sqrt{\sum_{t}^{T}(X_t-\overline{X})^2} \sqrt{\sum_{t}^{T}(Y_t-\overline{Y})^2}}
\end{equation*}

The range of correlation coefficient is between $1.0$ and $-1.0$. A value close to $1.0$ indicates a strong correlation between two variables and can express variable dependency. For instance, if the coefficient of spot placement and interruption-free score is close to $1.0$, we can assume that the two variables contain similar information. Meanwhile, a value close to $-1.0$ indicates a strong inverse correlation. A coefficient value close to $0$ indicates that two variables have no correlation and are more likely to be independent. Intuitively, the spot placement score and the interruption-free score should have a strong correlation because a higher spot placement score implies a higher likelihood of spot request fulfillment and a higher interruption-free score also implies a lower probability of spot interruption. Meanwhile, the two values are expected to have a strong negative correlation with the spot price because a higher spot price can be an indication of spot instance shortage.

\begin{figure}[t]
    \centering
    \includegraphics[width=0.42\textwidth]{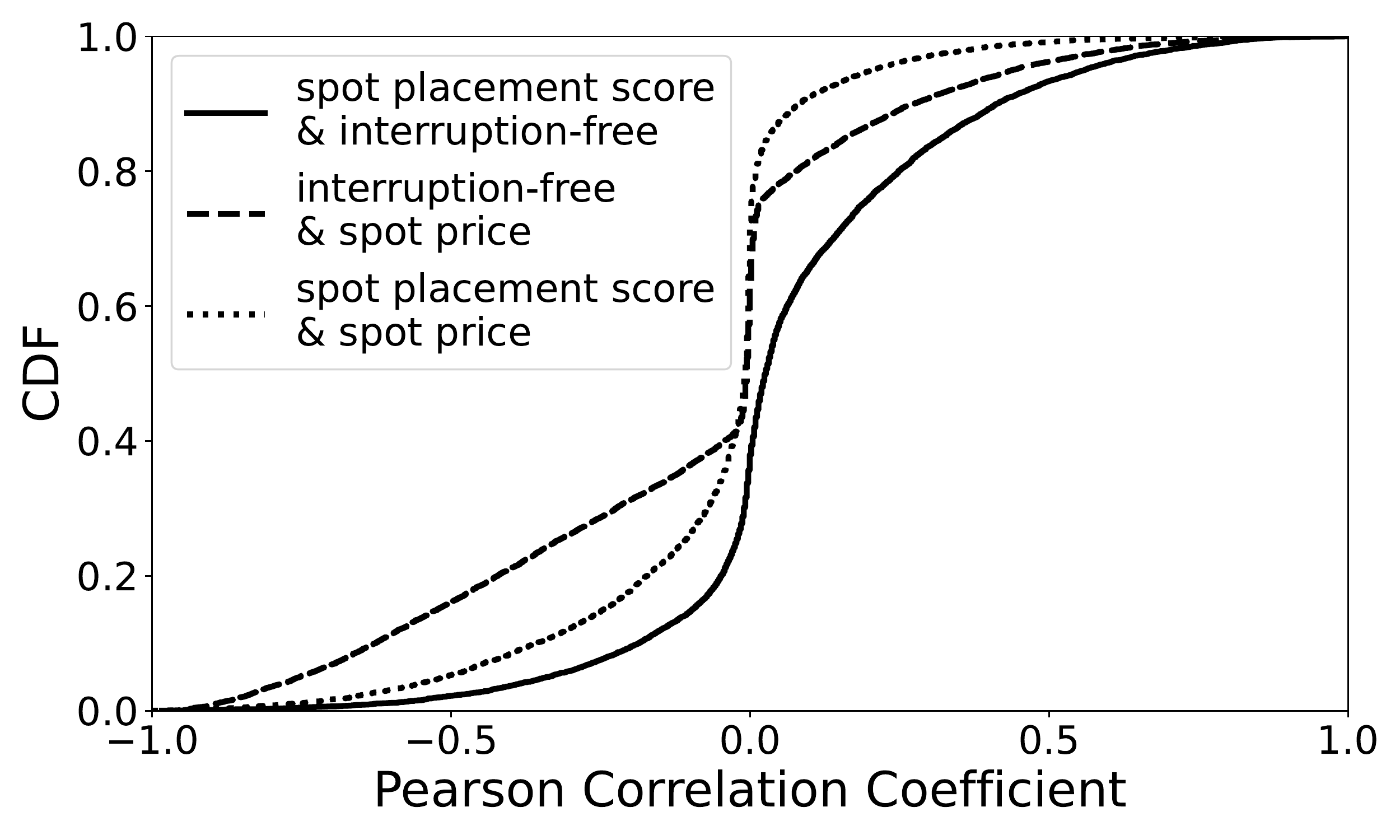}
    \caption{CDF of the Pearson correlation coefficient of any two combinations of the spot placement score, interruption-free score, and spot price.}
    \label{fig:correlation-coefficient-cdf}
\end{figure}

Figure~\ref{fig:correlation-coefficient-cdf} shows the Cumulative Distribution Function (CDF) of the Pearson correlation coefficient of the combination of any two variables. The horizontal axis presents the correlation coefficient values, and the vertical axis expresses the distribution. The solid line depicts the correlation coefficients between the spot placement score and the interruption-free score, the dashed line depicts the correlation coefficients between the interruption-free score and the spot price, and the dotted line depicts the correlation coefficients between the spot placement score and the spot price. As shown in the figures, most correlation coefficient values are located near $0.0$, which implies that the combination of any two spot datasets has neither strongly positive nor negative correlations. In the distribution, it is noticeable that the correlation coefficients that include the spot price have a much higher density around $0.0$, implying that the spot price dataset might have little information regarding the spot instance availability compared to the other two publicly announced spot instance datasets. This observation confirms what Irwin et al. discovered~\cite{spot-price-policy-change-2017-irwin} after the spot instance operation policy change in 2017. In addition, the correlation coefficients between the spot placement score and the interruption-free score are also very low. For $62.57\%$ of cases, the absolute coefficient value is lower than $0.25$, and $87.64\%$ of cases have correlation coefficients lower than $0.5$.

\begin{figure}[t]
    \centering
    \includegraphics[width=0.45\textwidth]{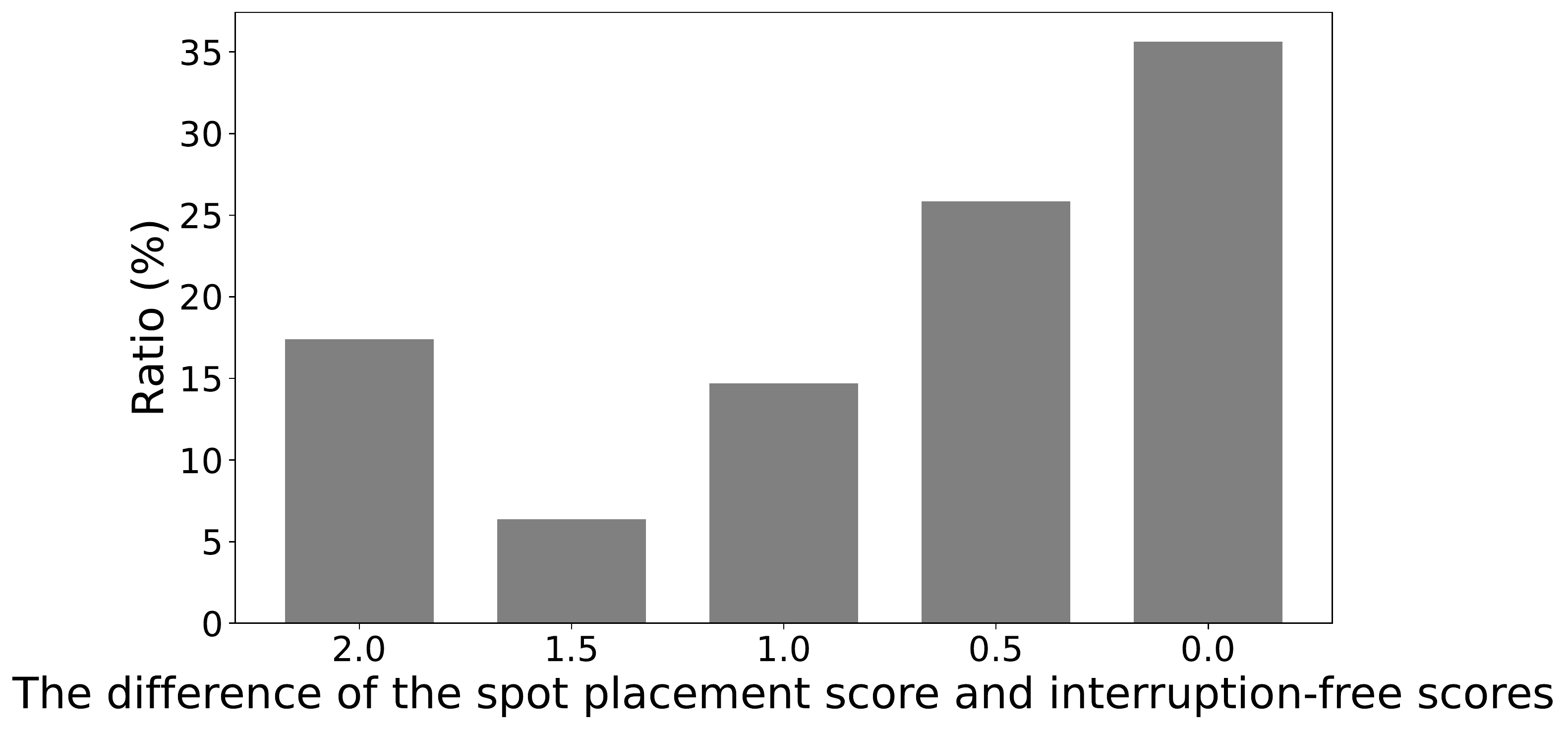}
    \caption{The distribution of score difference between the spot placement score and interruption-free score}
    \label{fig:sps-if-value-difference}
\end{figure}

The discrepancy among spot instance datasets can confuse users when the publicly announced datasets present contradicting information, such as a high spot placement score ($3.0$) with a low interruption-free score ($1.0$). To detect the extent to which the spot placement score and interruption-free score differs, we count the difference in the two scores at any given time and show the difference using a histogram in Figure~\ref{fig:sps-if-value-difference}. The horizontal axis shows the absolute score difference between the two datasets. The maximum and minimum of the scores are $3.0$ and $1.0$, respectively, and the step of the interruption-free score is $0.5$. Thus, the maximum difference value is $2.0$, which is a complete contradiction, and the minimum difference is $0.0$, which means the two score values are the same. The vertical axis displays the percentile unit ratio of each score difference. As illustrated in the figure, the difference of $0.0$ accounts for most cases. There are, however, numerous cases with contradictory information. For example, for about $17.41\%$ of cases, the spot placement and interruption-free scores present the opposite meaning. Considering that the difference of $1.5$ is not a negligible, for $24\%$ of cases, spot users might be confused about which datasets to follow for optimal spot usage.

\begin{figure}[t]
    \centering
    \includegraphics[width=0.45\textwidth]{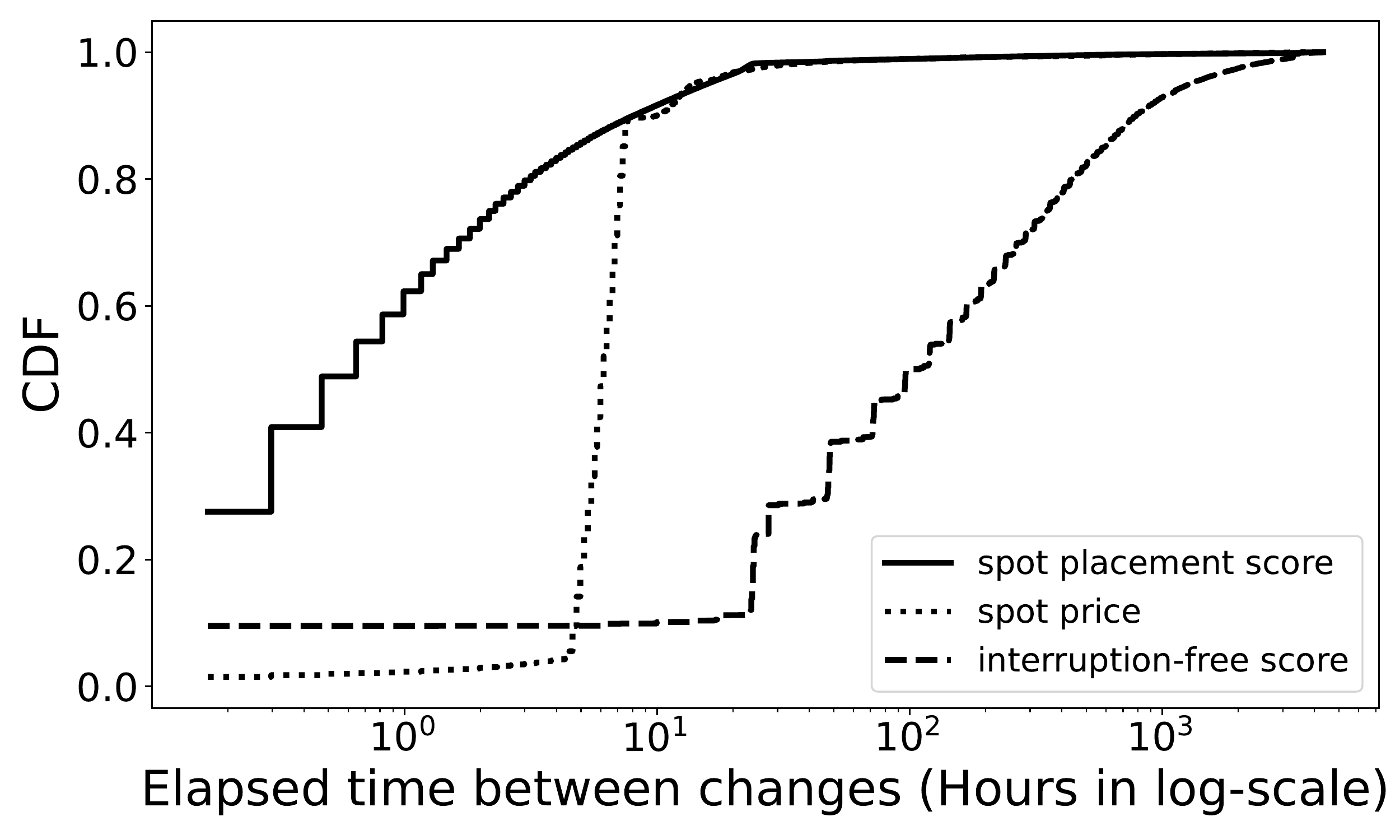}
    \caption{The distribution of the frequency of the value changes for the spot placement score, interruption-free score, and the savings over on-demand price.}
    \label{fig:change-frequency-cdf}
\end{figure}

To understand how often the spot dataset changes, Figure~\ref{fig:change-frequency-cdf} presents a CDF of elapsed time between updates. In the figure, the solid line represents the spot placement score, the dashed line represents the interruption-free score, and the dotted line represents the spot price. The horizontal axis of Figure~\ref{fig:change-frequency-cdf} expresses the elapsed time (hours) between update events in a log scale, with a lower the value on the horizontal axis indicating a more frequently updated variable. The spot placement score is updated the most frequently, while the interruption-free score is updated the least frequently. The interruption-free scores' low value change frequency is consistent with its score calculation policy, which uses the interruption ratio observed over the previous month. The frequent update of the spot placement score can be an indication of timely information that can reflect the success of the spot instance request well.

\textbf{Key findings}: The spot instance price, availability, and interruption-ratio do not show strong correlations, and they represent contradicting information quite often.

\subsection{Fulfilment and Interruption Behavior}\label{sec:real-world-exp}
To decide which spot dataset is a more reliable source of information, especially when they are contradicting, we conducted experiments to determine how well different spot datasets represent real-world spot instance behavior by measuring the spot instance fulfillment and interruption ratio of various instance types. The aim of the experiments was to identify how different spot placement and interruption-free scores affect spot instance availability. We decided to omit the impact from the spot price because it is known to no longer be a valid indication of spot instance availability~\cite{spot-price-change-2017, spot-instance-interrupt-check-cloud-2018, spot-price-policy-change-2017-irwin}.

In the experiments, we categorized the spot placement scores and interruption-free scores into \emph{\textbf{H}igh}, \emph{\textbf{M}edium}, and \emph{\textbf{L}ow,} whose values were $3.0, 2.0,$ and $1.0$, respectively. Then, we sampled instance type and availability zone, which belonged to one of the \emph{H-H, H-L, L-H, M-M}, and \emph{L-L} combinations where the character indicated the spot placement score and interruption-free score in order. The number of available instances in each combination differed, and we performed stratified under-sampling with the lowest number of available cases, which was the \emph{L-H} combination. With the stratified sampling~\cite{stratified-sampling}, we tried to distribute the instance type and availability zone uniformly across all the candidates. The pure random sampling resulted in a biased result to popular instance types and regions. In total, we generated $503$ experimental cases. Smaller and less expensive instance types were preferred where applicable to keep the experimental cost within our budget. For all the experimental cases, we issued a single spot instance request after setting the bid price the same as the on-demand price~\cite{how-not-to-bid-cloud} and recorded the request status every five seconds. In a spot request, we specified the \emph{persistent} parameter so that an interrupted instance was requested again soon after an interruption event. Each experiment scenario was conducted for $24$ hours.

\begin{table}[t]
    \centering
    \begin{tabularx}{0.45\textwidth}{|c||X|X|}
    \hline
        \makecell{Metric} & \makecell{Not-Fulfilled}   & \makecell{Interrupted}  \\ \hline
              
        H-H & \makecell{0\%}        & \makecell{14.71\%} \\
        H-L & \makecell{0\%}        & \makecell{40.52\%} \\
        M-M & \makecell{25.49\%}    & \makecell{39.22\%} \\
        L-H & \makecell{58.18\%}    & \makecell{30.91\%} \\
        L-L & \makecell{45.61\%}    & \makecell{45.61\%} \\ \hline
    \end{tabularx}
    \caption{The percentage of not-fulfilled and interrupted spot requests for different dataset category.}
    \label{table:non-fulfilled-interrupted-rate}
\end{table}

Table~\ref{table:non-fulfilled-interrupted-rate} presents the rates of cases that were not fulfilled and interrupted cases. For the \emph{Not-Fulfilled}, we counted cases that did not become fulfilled at all in the $24$-hour experiment. For the \emph{Interrupted}, we counted cases that were interrupted at least once during the experiment. It is noticeable that when the spot placement score was high, all the requests were fulfilled in the experiments. When both the spot placement and the interruption-free scores were high, $14.71\%$ of cases were interrupted at least once. When either the spot placement or the interruption-free score was medium or low, the interruption ratio skyrocketed to $45.61\%$ at most. It is worth noting that a low spot placement score is an indicator of fulfillment failure.

Overall, the success of fulfillment can be solely predicted by considering the spot placement score, which should reflect the most up-to-date resource availability information. This concurs with the score update frequency presented in Figure~\ref{fig:change-frequency-cdf}, which represented the shortest update frequency of the spot placement score. When predicting the interruption probability, it is more appropriate to consider both the spot placement and interruption-free scores where both values should score high.

\begin{figure}[t]
    \centering
    \subfloat[Latency until spot requests are fulfilled (shorter is better)]{
        \includegraphics[width=0.45\textwidth]{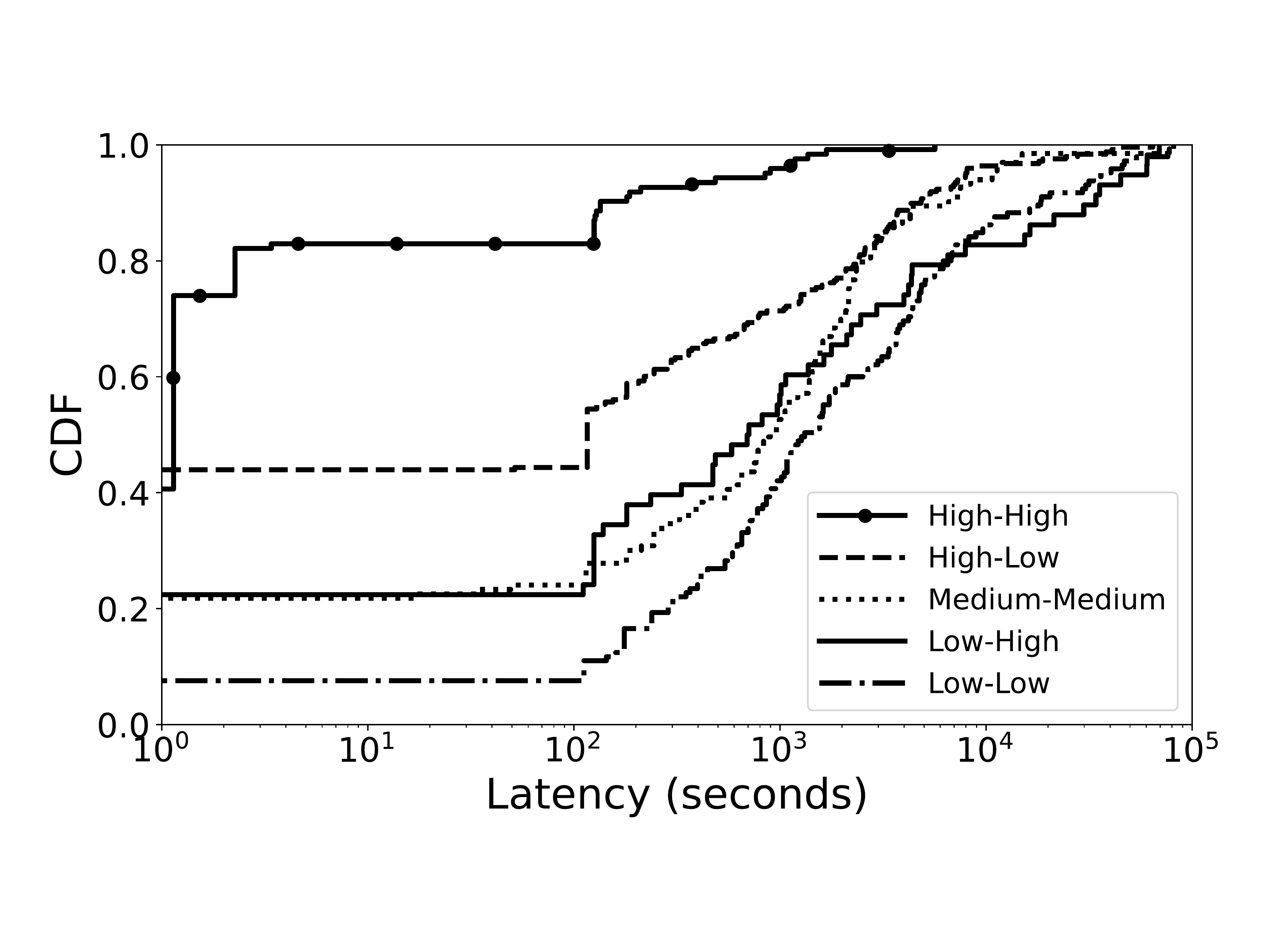}
        \label{fig:time-to-fulfill}
    }\\
    \subfloat[Time until an interruption event happens (longer is better)]{
        \includegraphics[width=0.45\textwidth]{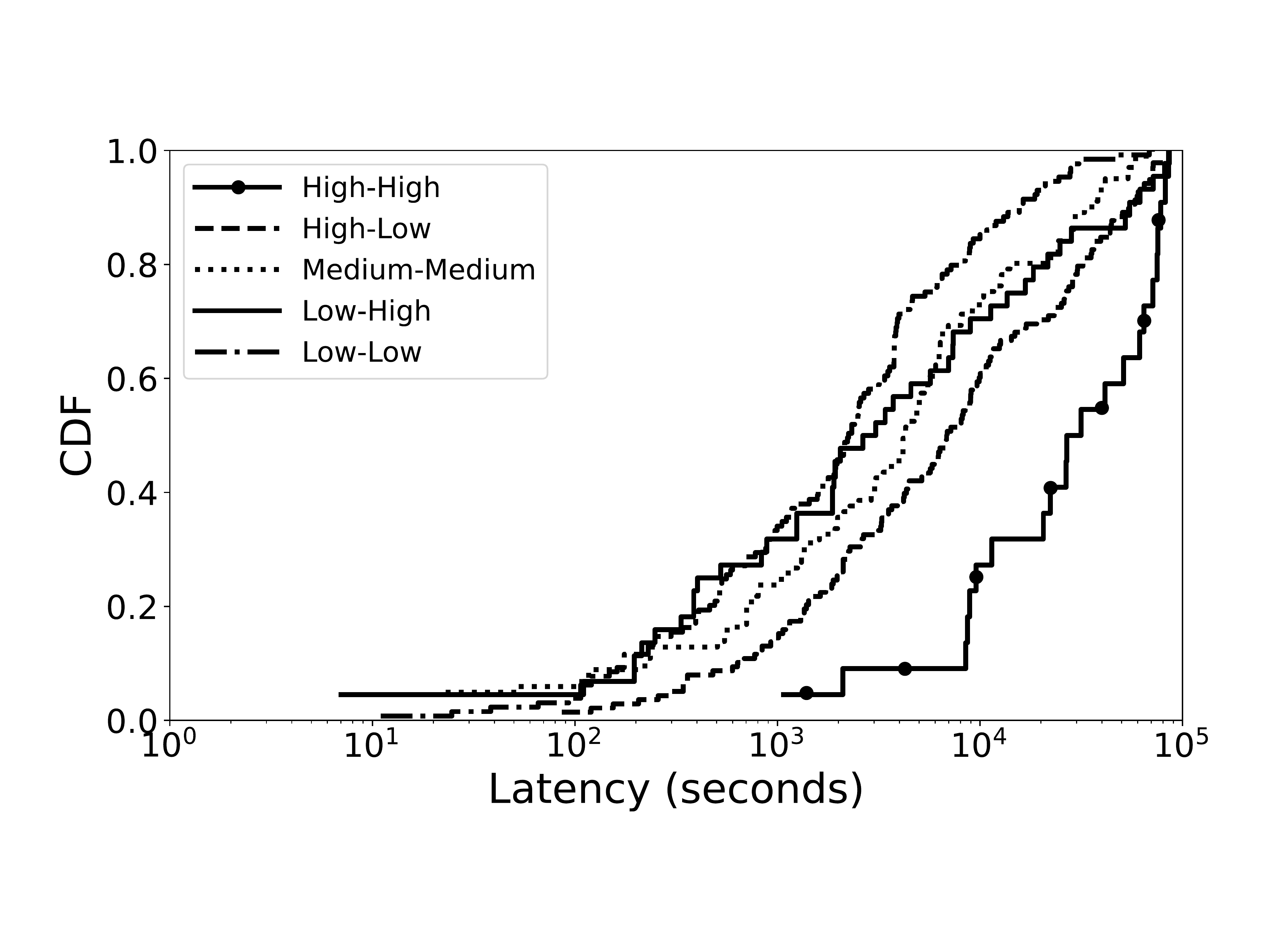}
        \label{fig:time-till-interruption}
\textbf{}    }
    \caption{The CDF of latency distribution categorized by the spot placement score and interruption-free score}
    \label{fig:spot-checker-time-cdf}
\end{figure}

To analyze how different spot placement and interruption-free scores impact the behavior of fulfillment and interruption, Figure~\ref{fig:spot-checker-time-cdf} presents the elapsed time from a spot request submission until it is fulfilled (Figure~\ref{fig:time-to-fulfill}) and the elapsed time from the fulfillment until an interruption event happens (Figure~\ref{fig:time-till-interruption}). Both figures are represented in a CDF format whose distribution is expressed on the vertical axis. The horizontal axis shows the elapsed time in seconds using the log scale. In each figure, we present distinct distributions after categorizing the spot placement and interruption-free scores into high, medium, and low. In Figure~\ref{fig:time-to-fulfill}, when both scores are high, about $28.07\%$ of requests are fulfilled within one second, and over $90\%$ of requests are fulfilled within $135$ seconds. When both scores are low, it takes the longest to fulfill spot requests with a median value of $1322$ seconds. When the two scores are contradictory, the higher spot placement score results in the faster fulfillment. 

Figure~\ref{fig:time-till-interruption} shows the time until a fulfilled spot instance becomes interrupted, and a larger value indicates higher availability. Similar to the time until fulfillment, the highest availability is observed when both scores are high, and the lowest availability is observed when both scores are low. The interruption ratio, which is presented in Table~\ref{table:non-fulfilled-interrupted-rate}, shows a similar value when the spot placement and interruption-free scores show distinct values. However, the spot instance running time shows noticeable differences where the median running time when the spot placement score is high (\emph{High-Low} case) is $6872$ seconds, while that of when the interruption-free score is high(\emph{Low-High}) is $2859$ seconds.

\textbf{Key findings}: When both spot placement and interruption-free scores are high, it is the most reliable, but when the two scores imply different information, the spot placement score should take precedence.

\begin{table}
    \centering
    \begin{tabularx}{0.48\textwidth}{|c||X|X|X||X|}
    \hline
        \makecell{Metric} & \makecell{IF}   & \makecell{SPS} & \makecell{Cost Save} & \makecell{RF}  \\ \hline
        Accuracy    & \makecell{0.45} & \makecell{0.64} & \makecell{0.39} & \makecell{\bf0.73} \\
        F1-score     & \makecell{0.43} & \makecell{0.58} & \makecell{0.28} & \makecell{\bf0.73} \\ \hline
    \end{tabularx}
    \caption{Spot instance status prediction method performance comparison. Using the historical dataset provided by the proposed work (RF) shows the best performance}
    \label{table:prediction-model-eval}
\end{table}

\subsection{Usefulness of Historical Spot Instance Dataset}\label{sec:sps-ml-model}
One of the main contributions of the proposed spot instance data archive service is the offering of an historical dataset. Using the historical dataset, we can take advantages of abundant prior information by creating a machine learning model to predict spot instance interruption events. To present the practical advantage, we built a simple random forest~\cite{random-forest} model using a Python Scikit-Learn package~\cite{scikit-learn} with default parameters without tuning. For training and testing, we used the real-world spot instance interruption observation experimental results in Section~\ref{sec:real-world-exp}. Using the dataset, we defined a classification problem whose target classes were \emph{NoInterrupt, Interrupted}, and {NoFulfill}. When building the simple prediction model, we used the historical spot placement score and interruption-free score of the preceding month as features. To compare the prediction accuracy of the machine learning model that uses the historical dataset with that of a simple heuristic that does not use historical information, we used three distinct heuristics that only used the current spot placement score, interruption-free score, and cost savings. When using the spot placement score to predict instance interruption, the score being $3.0$ at the time a spot instance request was made was classified as \emph{NoInterrupt}, being $2.0$ was \emph{Interrupted}, and being $1.0$ was \emph{NoFulfill}. We set the thresholds for interruption-free score and cost savings empirically after numerous trials.

Table~\ref{table:prediction-model-eval} shows the spot instance status prediction performance. In the table, the IF (interruption-free), SPS (spot placement score), and Cost Save columns represent prediction heuristics that reference only the current information. The three heuristics are implementable using the current spot instance datasets. The last column of RF represents a prediction model that can use abundant historical spot dataset information, which is available through the proposed data archive service. We compare the different methods using accuracy and F1-score, which reflects both precision and recall~\cite{ml-evaluation-metric}. As shown in the table, the simple machine learning model that that uses historical information from this proposed work outperforms a heuristic that references only the current information. Considering that the prediction model in this evaluation is fairly simple, the research community could likely propose a more elaborate and fine-tuned model using the proposed service.

\textbf{Key findings} : Using the historical spot dataset information provided by the proposed data archive service has high potential to predict future spot instance availability, which is vital to enhancing optimal cloud usage.

\section{Related Work}
\textbf{Using spot instance price dataset}: The spot instance price change history dataset has been widely used for various purposes. Uncovering statistical characteristics of the spot price change~\cite{draft-spot-instance-guarantee-from-spot-price, deconstructing-spot-instance, spot-analysis-javadi, stat-analysis-spot-price, spot-instance-analysis, spot-price-by-location, spot-instance-for-hpc} allows spot instance users to estimate the resource availability and cost savings when using spot instances. However, the spot instance operation policy change in 2017~\cite{new-spot-price} made the previous analysis work obsolete, and the price change data itself does not provide information that is as rich as it was before~\cite{spot-price-policy-change-2017-irwin, spot-price-change-2017}. Irwin et al.~\cite{spot-price-policy-change-2017-irwin} thoroughly discussed the advantages and disadvantages of the spot price policy change while suggesting a direction for the spot instance evolution. New spot instance datasets have been released since the policy change, but they have received comparatively little attention. To the best of the authors' knowledge, this is the first study to empirically examine spot instance availability information using a composite of multiple spot instance datasets.

\textbf{Modeling spot instance availability}: Kadupitige et al.~\cite{google-cloud-empirical-preemption} proposed a statistical model to represent the constrained spot instance preemption through empirical experiments using Google Cloud Spot VMs. The Google cloud does not provide transient instance resource availability information as AWS does, and Kadupitige et al. tried to build a statistical interruption model. Meanwhile, AWS keeps opening new spot instance datasets, and we attempted to analyze the characteristics and verify how the dataset resembles real-world spot instance behavior. Pham et al.~\cite{spot-instance-interrupt-check-cloud-2018} empirically analyzed the spot instance availability with the interruption frequency dataset provided by AWS. As presented in Section~\ref{sec:real-world-exp}, the spot instance fulfillment and interruption probability can be better modeled by combining the spot instance availability and interruption frequency datasets, and providing the historical information can open up many opportunities for further research.

\textbf{Usefulness of spot instance availability information}: By referencing availability information from spot instance datasets, different types of applications can prepare a plan to react to spot instance interruptions. Son et al. proposed DeepSpotCloud~\cite{deepspotcloud} to run DNN training tasks using GPU spot instances located globally. Big data analysis tasks are generally conducted with Hadoop~\cite{mapreduce, hadoop}, and SeeSpotRun~\cite{see-spot-run} proposed running Hadoop MapReduce tasks using spot instances. Flint~\cite{flint} and Tr-Spark~\cite{tr-spark} proposed a system to run a distributed big data processing engine, Apache Spark~\cite{spark}, using spot instances. Using spot instances, online web services~\cite{spot-for-online-service, spotweb}, batch processing jobs~\cite{spot-batch, spoton}, and parallel processing of independent tasks~\cite{autobot-bot-using-spot} while mitigating the straggler effect due to transient servers~\cite{spot-straggler-mitigation} have been proposed. ExoSphere~\cite{exosphere} proposed a portfolio modeling for applications with different levels of interruption-tolerance and cost reduction expectations. The findings in this paper are complement the aforementioned work that heavily used spot instances because the historical spot placement availability information can help to improve the accuracy of the spot instance availability prediction, which is no longer possible with the spot price dataset.

\section{Extending Service for Various Cloud Vendors}
The functionalities of the proposed data archive service and this paper are mainly focused on AWS Spot instances. One of our actively ongoing work is to add spot instance datasets of various cloud service vendors. Microsoft Azure provides current spot instance price information via the API and web portal service. Azure provides spot instance availability and interruption ratio information only from its web portal. Google Cloud provides the current spot instance price only from its web portal. Considering that most cloud vendors provide only current information with a limited data access method, accessing the spot dataset through the proposed data archive service will greatly enhance cloud usage from various vendors with numerous research opportunities.

To expand SpotLake service to provide spot instance dataset from various cloud vendors, we have to develop a data scheme for distinct datasets from multiple vendors as they provide a different set of spot instance information. Despite the distinct information, a global key would help more complex analyses such as composite spot instance analysis over the multiple cloud vendors. For instance, we are currently developing data collection for multiple vendors using the timestamp as a global key, and it helps to understand temporal characteristics of spot instance availability pattern of different cloud providers. Other than the timestamp, adding more global keys such as hardware details are beneficial to analyze and compare the spot instance characteristics from various aspects and help to find optimal spot resources for diverse workloads. Comparing spot instances of multiple vendors in a single place can provide a great opportunity for optimal resource usage and serve to enhance multi-cloud service, which is deemed to be the direction of cloud computing evolution~\cite{sky-above-clouds, optimal-multi-cloud-configuration}.

\section{Conclusion}
The spot instance provided a more affordable way to use cloud instances, and the spot price history dataset was a valuable source of information to predict spot instance availability. However, as the price change became less volatile, most prior work that relied on the spot price datasets became obsolete. To help users to better utilize various spot instance datasets, we built a data archive service that provides historical and current information of spot instance availability, interruption ratio, and price datasets that are challenging to assemble due to various data access constraints. Using the collected data, we performed a thorough empirical analysis of the spot instance behavior to determine how different spot datasets represent the instance availability. We also presented the applicability of the proposed historical spot instance dataset archive by applying a simple machine learning model that enhances the prediction accuracy of future spot instance availability. The proposed system is currently publicly available, and we are sure that it will enhance the system research in the field of optimal cloud resource usage with spot instances. 

\section*{Acknowledgments}
We would like to thank anonymous reviewers and program chairs for their insightful comments and feedback. Special thanks to Chaelim Heo, who helped the SpotLake web service front-end implementation. This work is supported by the National Research Foundation (NRF) Grant funded by the Korean Government (MSIP) (No. NRF-2022R1A5A7000765 and NRF-2020R1A2C1102544), AWS Cloud Credits for Research program, and the SW Star Lab (RS-2022-00144309) of IITP.

\bibliographystyle{IEEEtranS}
\bibliography{spot-price-score}

\clearpage

\appendix
\section{Artifact Appendix}

\subsection{Abstract}
This artifact describes steps to reproduce figures and tables on this paper with necessary datasets. We provide preprocessed data that is downsized from raw data and source code to reproduce figures and values. To access the raw dataset with further information, you can check our official SpotLake web service. 

\subsection{Artifact check-list}
{\small
\begin{itemize}
  \item {\bf Run-time environment: Python 3.10}
  \item {\bf Necessary packages: Scipy, Numpy, Pandas, Scikit-learn, Matplotlib, Seaborn}
  \item {\bf Hardware: AWS EC2 t3 micro instance}
  \item {\bf Output: Figures(PDF), values of tables}
  \item {\bf How much disk space required?: Under 100MB}
  \item {\bf How much time is needed to generate figures and tables?: 1 minute}
  \item {\bf Publicly available?: Yes}
  \item {\bf Code licenses (if publicly available)?: Apache License 2.0}
  \item {\bf Archived (provide DOI)?: Yes (\url{http://doi.org/10.5281/zenodo.7084392})}
\end{itemize}
}

\subsection{Description}

\subsubsection{How to access}
You can access to preprocessed data and source code on a public archive platform
\begin{itemize}
    \item \url{http://doi.org/10.5281/zenodo.7084392}
\end{itemize}

The main contents of the artifacts in the public archive contain followings:
\begin{itemize}
    \item Python source code to generate from Figure 3 to 11
    \item Python source code to generate Table 1, 2, and 3
    \item Preprocessed data used in the source code
    \item Generated figures using the source code
\end{itemize}

\subsection{Installation}
These are the installation steps to prepare necessary packages to generate figures and values of tables.

\begin{enumerate}
    \item Set up your environment with \emph{Python 3.10}
    \item Download and unzip artifact to your environment
    \item Move to \emph{codes} directory
    \item Install python packages using \emph{pip install -r requirements.txt}
\end{enumerate}

\subsection{Evaluation and expected results}
After the necessary library installation, you can generate figures and values by running python codes that are included in \emph{codes} directory in the artifact. After running a figure generation source code, such as \emph{figure03-a.py}, you can check the generated PDF format figures on \emph{figures} directory. The table value generation source codes, such as \emph{table01.py}, does not generate any file, but it prints only values presented in the tables. The dataset that is referenced from the source code is stored in the \emph{data} directory with a corresponding figure and table indexes in a \emph{pickle} file format.

\subsection{Notes}
If you want to access the collected raw dataset, detailed collection methodology, and system implementation source code, you can visit our official SpotLake web service in \url{https://spotlake.ddps.cloud}. In the website, you can access the following information.

\begin{itemize}
    \item Historical spot availability dataset
    \item SpotLake data collection source code
    \item SpotLake system implementation source code
\end{itemize}

\end{document}